\documentclass[letterpaper, 10 pt, conference]{ieeeconf}  % Comment this line out if you need a4paper

\IEEEoverridecommandlockouts                              % This command is only needed if 
\usepackage{epsfig} % for postscript graphics files
\usepackage{amsmath}
\usepackage{xcolor}
\usepackage{comment}
\newtheorem{theorem}{Hypothesis}
\newtheorem{teorema}{Theorem}
\usepackage{algorithm}
\usepackage[noend]{algpseudocode}

\title{\LARGE \bf Model Predictive Control with adaptive resilience for Denial-of-Service Attacks mitigation on a Regulated Dam}

\author{Raffaele G. Cestari$^{1}$, Stefano Longari$^{1}$, Stefano Zanero$^{1}$, Simone Formentin$^{1}$% <-this % stops a space
\thanks{$^{1}$Department of Electronics, Information, and Bioengineering, Politecnico di Milano,
Milano, Italy. Email to: {\tt\small raffaelegiuseppe.cestari@polimi.it}.}
\thanks{This paper is partially supported by FAIR (Future Artificial Intelligence Research) project, funded by the NextGenerationEU program within the PNRR-PE-AI scheme (M4C2, Investment 1.3, Line on Artificial Intelligence), by the Italian Ministry of Enterprises and Made in Italy in the framework of the project 4DDS (4D Drone Swarms) under grant no. F/310097/01-04/X56 and by the PRIN PNRR project  P2022NB77E “A data-driven cooperative framework for the management of distributed energy and water resources” (CUP: D53D23016100001), funded by the NextGeneration EU program.} 
}

\begin{document}

\maketitle
\thispagestyle{empty}
\pagestyle{empty}

%%%%%%%%%%%%%%%%%%%%%%%%%%%%%%%%%%%%%%%%%%%%%%%%%%%%%%%%%%%%%%%%%%%%%%%%%%%%%%%%
\begin{abstract}
In recent years, SCADA (Supervisory Control and Data Acquisition) systems have increasingly become the target of cyber attacks. SCADAs are no longer isolated, as web-based applications expose strategic infrastructures to the outside world connection. In a cyber-warfare context, we propose a Model Predictive Control (MPC) architecture with adaptive resilience, capable of guaranteeing control performance in normal operating conditions and driving towards resilience against DoS (controller-actuator) attacks when needed. Since the attackers' goal is typically to maximize the system damage, we assume they solve an adversarial optimal control problem. An adaptive resilience factor is then designed as a function of the intensity function of a Hawkes process, a point process model estimating the occurrence of random events in time, trained on a moving window to estimate the return time of the next attack. We demonstrate the resulting MPC strategy's effectiveness in 2 attack scenarios on a real system with actual data, the regulated Olginate dam of Lake Como.
\end{abstract}

%%%%%%%%%%%%%%%%%%%%%%%%%%%%%%%%%%%%%%%%%%%%%%%%%%%%%%%%%%%%%%%%%%%%%%%%%%%%%%%%
\section{INTRODUCTION}

In recent years, the number of attacks on SCADA (Supervisory Control and Data Acquisition) systems has soared, leading to 2021 as the year in which, for the first time, the manufacturing and critical infrastructure industries suffered the most tremendous economic and structural damage, according to the IBM 2021 report. The most famous is the Stuxnet malware (2010), which targeted a Nuclear Power Plant in Iran. Night Dragon (oil and gas, 2010), Shamoon (oil and gas, 2012), New York dam (2013), Ukraine power grid (2015-2016), up to the most recent ones (OMRON, Schneider PLCs, 2022) are just some of the malwares and targets of the last few years. Citing Drias et al.~\cite{drias2015analysis}, \textit{avoiding attack is quite impossible. Industrial networks should perform profiling, network traffic monitoring, and attack detections. Industrial network infrastructure must be able to ensure control operations under attacks due to the criticality of the systems that they operate.} As they point out, it is impossible to prevent the attacks. This depends on the disparity between the resources needed by attackers and defenders: developing malware is relatively simple and easy to deploy, and conversely, repairing the damage caused by it is expensive in terms of money and time. For this reason, it is crucial to study the development of control schemes from a performance maximization perspective and consider the possibility that they are victims of cyber threats. 
SCADA systems often constitute national and international safety critical infrustructures such as power grids and dams. For this reason, they represents privileged targets of cyber-warfare and might cause enormous social and economic damages on national and international scale (see \cite{gandhi2011dimensions}, \cite{rudner2013cyber}, \cite{lis2019cyberattacks}).

The primary attacks in the literature are \textit{Deception}, \textit{False Data injection}, \textit{Eavesdropping}, and \textit{Denial of Service (DoS)}. See \cite{irmak2018overview}.
The first two consist of the injection of corrupted data into the system. Deception, more insidious than false data injection, introduces corrupted data with continuity properties, making detecting the attack more complex. See \cite{du2019malicious}. Eavesdropping consists of violating the confidentiality of the communication, sniffing packets on the Local Area Network (LAN), or intercepting wireless transmissions. 

DoS attacks are the most common (an easy to deploy) and lead to the interruption of communications between the controller and actuator (or analogously between controller and sensors) and, in the worst case, to the application of malicious control actions. See \cite{markovic2013analysis}, \cite{kalluri2016simulation}. Given its ease of development and applicability, in this work we focus on developing a model predictive control architecture capable of mitigating the impact of DoS attacks on the communication between controller and actuator. 

In the literature, several formulations of Resilient MPC aim to respond through an appropriate control scheme design to one or more of these threats. The authors of \cite{liu2020mpc} propose an observer that, combined with MPC, is capable of detecting deception attacks. In \cite{habibi2021secure} they propose an artificial neural network (ANN) based MPC to identify occurrence of false data injection attacks in DC microgrids. The work in \cite{sun2019resilient} proposes a resilient MPC architecture against DoS attacks jamming the communication between controller and actuator. Their main idea is to store actuator-side the whole vector of optimal control actions computed by the MPC over the prediction horizon in such a way that, even if the data transmission between controller and actuator is compromised, the actuator is still capable of executing optimal control actions. 

Our problem statement takes a stronger assumption: we assume that the actuator side is hijacked by the attacker. Communication between controller and actuator is compromised and the actuator is no longer in control, hence the proposed solution of \cite{sun2019resilient} would not be feasible. 

Suppose that the attacker can access the actuator at any time and tamper with the optimal control action for that instant. Our architecture will mitigate the damage caused by fiddling with the optimal control. In particular, the proposed solution employs two MPC controllers which simultaneously solve two different optimization problems. The master controller, hereafter called ResMPC, is responsible for calculating the optimal control action applied to the system. This controller determines the optimum by balancing the performances obtainable on the prediction horizon by including a "regularizing" term that weighs the distance of its optimal control action from that provided by the second controller, from now on called SafeMPC, which instead assumes that from the next instant and after that, the actuator will be out of control and applying the optimal control action on the system will no longer be possible.

From this perspective, it is clear how impactful it is to correctly balance performance tracking with respect to the safe control action that assumes the presence of an attack in the immediate future. Choosing the conservative path in the absence of attacks will lead to a loss of performance, while in the presence of frequent attacks, pursuing only the quality of the control action on the prediction horizon will cause a blind approach to the onset of attacks with a consequent damage overhead compared to the precautionary solution.

For this reason, we propose an online adaptive solution that can estimate the onset of attacks in the near future and increase or decrease the resilience factor responsible for the balance between performance and safety.

This is possible thanks to Hawkes processes (\cite{laub2015hawkes}), a class of point processes that allows us to estimate the return time of an event whose occurrence is apparently random in time. Hawkes processes find wide use in the literature in all those applications where predicting events without uniform temporal sampling is necessary. Some examples are finance (\cite{cestari2023hawkes}, \cite{bacry2015hawkes}), neuroscience (\cite{reynaud2013inference}), seismology (\cite{kwon2023flexible}), and social media (\cite{rizoiu2017hawkes}). We have references also in the field of cybersecurity. The authors of \cite{bessy2021multivariate} used multivariate Hawkes processes to predict attack frequency.
In our architecture, at each simulation step, we train the Hawkes process on a fixed-length moving window (expanding at the beginning of the simulation as we assume that we have no data before that collected online) and calculate its intensity function as the return time of the attack varies. Since the intensity function measures the probability of event occurrence, it will increase as the expected return time decreases. By defining a fixed look-ahead horizon, we integrate the intensity function to correspondingly define the value of the resilience factor. In this way, we algebraically associate the intensity function with the resilience factor, representing the estimated attack condition in the near future.

We apply the proposed methodology on a SCADA system, linear with saturation, represented by the water release from the Olginate dam responsible for regulating the Lake Como basin, Lombardy, Italy. The system to be regulated (with application-related problems) and the "performance-oriented" reference MPC controller are described in \cite{cestari2022hourly} and \cite{cestari2023scenario}. The data used in the simulation are authentic and come from the dam's historical regulation archive. This case study is well representative of a critical infrastructure for the management of water resources in the Lombardy region, in which the lack of "resilience" to cyber attacks could lead to direct (flooding of roads, houses, commercial and industrial activities in the city of Como) and indirect (failure to release water to agricultural districts in periods of drought leads to significant economic losses) economic damage.
To focus exclusively on the attack impact, we assume the perfect knowledge of the system disturbances (e.g., the water inflow in the lake).

Given the application, we assume that we are working in a cyber warfare context in which the attacker is interested in undermining the territorial and infrastructural stability of the country. From this perspective, we assume that it is capable of exercising through a DoS attack the interruption of communications between the dam controller and the actuation of the release valve and that it takes its place, applying the policy of "maximum damage" that is, given the conditions (states) of the system, evaluate which control action could cause the greatest possible damage. To make this possible, we assume that the attacker has access, at every time instant, to the measured states of the system and that to determine the "optimal attack", it solves an optimization problem equal and opposite to the one solved, at the same instant, by the MPC controller, in order to maximize the damage caused by the malicious control action.

We evaluate performance on actual data of year 2005, comparing 3 control architectures:
\begin{itemize}
    \item MPC: the standard "performance-oriented" MPC, described in \cite{cestari2022hourly} and \cite{cestari2023scenario}.
    \item SafeMPC: the "safety-oriented" MPC, applying to the actual system exclusively the control action computed by the SafeMPC controller.
    \item  Adaptive Resilient MPC (ARMPC): the proposed architecture, adapting online resilience by balancing performance and safety given the expected return time of the attacks.
\end{itemize}

We compare the controllers in 2 attack scenarios:
\begin{itemize}
    \item A time-varying periodic attack pattern, in which the attacks happen periodically with changing period.
    \item A pseudo-random binary attack pattern, in which the attacks happen pseudo-randomly given the firing of a pseudo-random binary periodic triggering signal.
\end{itemize}

We show that adaptive resilience is mandatory for adaptation to changing attack conditions and that the proposed architecture, Adaptive Resilient MPC, outperforms the competitors in both the attack scenarios.

The main contributions of the paper are the following:
\begin{itemize}
    \item As far as we are aware, for the first time, we propose a model predictive control architecture capable of mitigating DoS attacks where the attacker overrides the optimal control action and replaces it with the optimal adversarial control action. 
    \item We propose an online adapting solution capable of predicting the attack occurrence and correspondingly increasing the safety level, mitigating damages, and, simultaneously, keeping a satisfactory performance tracking in periods of absence of attacks. 
    \item For the first time, as far as we are aware, we exploit Hawkes processes in the context of model predictive control to forecast cyber attack occurrence.
\end{itemize}

The remainder of the paper is as follows. In Section \ref{mathematical} we describe the Adaptive Resilient MPC architecture and the optimal attack problem. In particular, Section \ref{ResMPC} describes the optimal control problem solved by the main ResMPC controller. Section \ref{safeMPCSect} describes the optimal control problem solved by the auxiliary "safety" controller. Section \ref{OptimalAttacker} describes the adversary optimal control problem solved at each time step by the attacker. Section \ref{motivations} shows why an adapting resilience factor is needed. Section \ref{hawkes} describes the Hawkes process and the resilience factor computation. In Section \ref{results} we show the numerical simulations and results in the 2 attack scenarios. Section \ref{conclusions} summarizes the main contributions of the paper and the future work.

\section{Mathematical Formulation}
\label{mathematical}
\begin{figure*}[htbp!]
\centering
\includegraphics[width=0.75\textwidth]{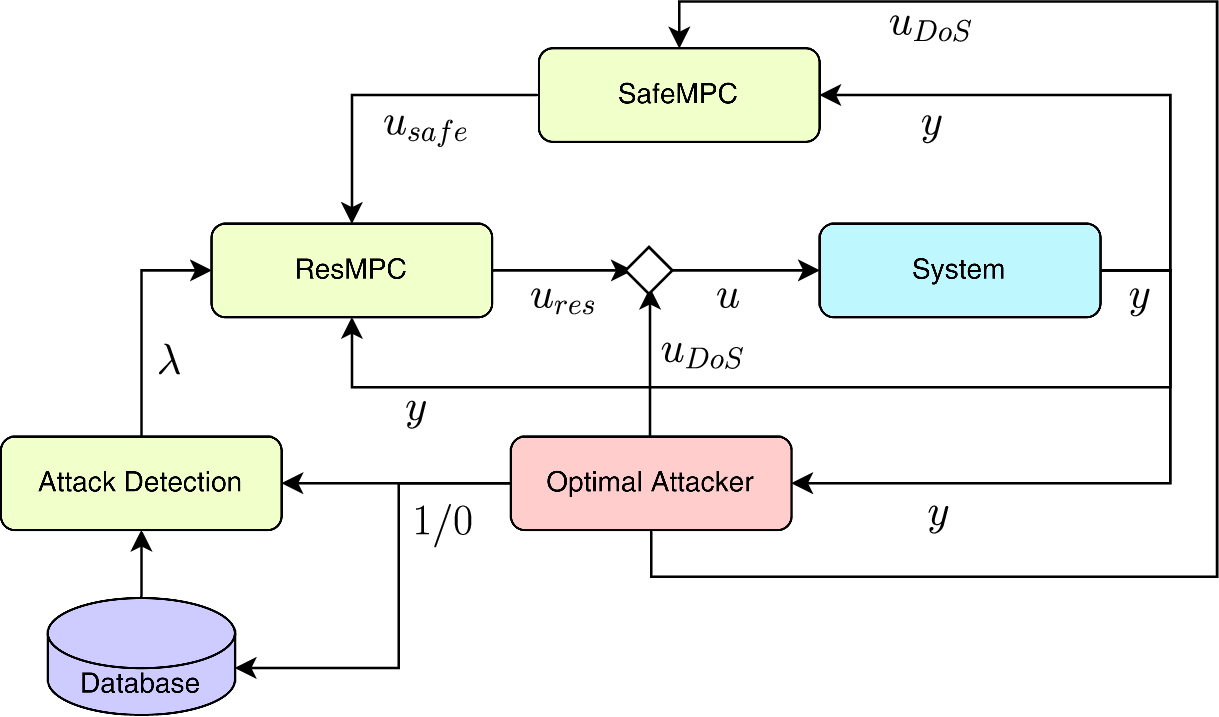}
\caption{Adaptive Resilient MPC Architecture}
\label{fig:scheme}
\end{figure*}

The Resilient MPC architecture consists in 2 MPC controllers. The main one, \textbf{ResMPC}, is responsible for computing the optimal control action $u_{res}$, at each time step, considering simultaneously control performance and safety. The auxiliary controller, \textbf{SafeMPC}, computes, at each time step, the safe control action $u_{safe}$ which is fed to the \textbf{ResMPC}. The attacker interrupts the communication between controller and actuator and replaces the optimal control action $u_{res}$ with its optimal malicious action $u_{DoS}$. Attack occurrence is saved as a binary variable in a database. The attack detection module provides the resilience factor $\lambda$ to the \textbf{ResMPC} after training on historical archive and current time attack occurrence. Figure \ref{fig:scheme} summarizes the Adaptive Resilient MPC architecture.

\subsection{Resilient MPC Control Problem}
\label{ResMPC}
\textbf{ResMPC} controller solves, at each time step, the following constrained, convex optimization problem:
\begin{subequations}
	\label{MPC}
	\begin{equation}
            \label{cost1}
		\operatorname{min}_{u}\{c_f||\textbf{s}-s_{max}||_2 + c_d||\textbf{s}-s_{min}||_2+c||\epsilon||_2 + \lambda ||\textbf{u}-u_{safe}||\}
	\end{equation}    
	\begin{equation}
        \label{massBalance1}
		s_{h+1} = s_h +3600\cdot 24(\hat{q_h} - u_h) 
	\end{equation}
	\begin{equation}
        \label{controlConstrainta}
		u_{min} \leq u_h \leq u_{max}
	\end{equation}
	\begin{equation}
        \label{controlConstraint2a}
		u_h \geq d_h +\epsilon_h \\
	\end{equation}
         \begin{equation}
         \label{derivativeConst1a}
		u_h - u_{h-1} \leq 240 \,\, m^3/s \,\,\,\, if \,\, l_h \leq 1.1 \,\, m \\
	\end{equation}
         \begin{equation}
         \label{derivativeConst2a}
		u_h - u_{h-1} \leq 360 \,\, m^3/s \,\,\,\, if \,\, l_h > 1.1 \,\, m \\
	\end{equation}
	\begin{equation}
			\textbf{s} \in R^H, \textbf{u} \in R^H, \epsilon \in R^H
	\end{equation}
        \begin{equation}
        \label{values}
			c_f = 10^{-4}, c_d = 10^{-3}, c=100
	\end{equation}
\end{subequations}
Equation \eqref{cost1} shows the minimization problem. $\textbf{u}$ $[m^3/s]$ is the control action (e.g., the water release going out from the lake) vector over the prediction horizon $H$. $c_f$, $c_d$ and $c$ are numerical coefficients tuned to have the same order of magnitude for the 3 objectives whose value is collected in Equations \eqref{values}, for further details, see \cite{cestari2022hourly}, \cite{cestari2023scenario}. $\textbf{s}$ $[m^3]$ is the lake volume vector. $s_{min}$ and $s_{max}$ are the minimum and maximum lake volumes. The control goal is, simultaneously, maintaining the lake volume between the minimum and maximum thresholds (dry and flood levels, first and second cost components respectively) and satisfying the agricultural water demand (third component). The fourth cost component is the key of resilience: we want to minimize the distance of the following optimal control action that will be applied to the system from the optimally computed safe control action $u_{safe}$ $[m^3/s]$. $\lambda$ is the resilience factor, increasing it leads the controller to safer decisions at the cost of losing performance quality. Equation \eqref{massBalance1} shows the daily mass balance equation for the lake volume. $\hat{q}_h$ is the predicted water inflow entering the lake (in this work we assume it coincides with the actual one) in the $h$ time-step of the prediction horizon. Equation \eqref{controlConstrainta} shows the control action constraints, minimum $u_{min}$ and maximum $u_{max}$ respectively. These constraints are non-linear functions of the lake level. For additional details see \cite{cestari2022hourly}. Equation \eqref{controlConstraint2a} shows an additional soft constraint on control action. Water release must be close to the agricultural water demand $d$. The constraint must be soft to include cases where the release is lower than the request (dry periods) and where it must be above the request (flood periods). This is ensured through the slack variable $\epsilon$, an additional penalty term in the cost function \eqref{cost1}. Equations \eqref{derivativeConst1a} and \eqref{derivativeConst2a} ensure that the variation of control action is always below the daily limits imposed by the Olginate dam operation. These boundaries depend on $l \,\, [m]$, lake level.

\subsection{Safe MPC Control Problem}
\label{safeMPCSect}
The safe controller solves, at each time step, a constrained convex optimization problem whose role is computing the next safe control action $u_{safe}$:
\begin{subequations}
	\label{SafeMPC}
	\begin{equation}
            \label{costSafe}
		\operatorname{min}_{u_{safe}}\{c_f||\textbf{s}-s_{max}||_2 + c_d||\textbf{s}-s_{min}||_2+c||\epsilon||_2\}
	\end{equation}    
	\begin{equation}
        \label{massBalance}
		s_{h+1} = s_h +3600\cdot 24(\hat{q_h} - u_h) 
	\end{equation}
	\begin{equation}
        \label{controlConstraintSafe}
		u_{min} \leq u_{safe} \leq u_{max}
	\end{equation}
	\begin{equation}
        \label{controlConstraint2}
		u_{safe} \geq d_h +\epsilon_h \\
	\end{equation}
         \begin{equation}
         \label{derivativeConst1}
		u_{safe} - u_{t-1} \leq 240 \,\, m^3/s \,\,\,\, if \,\, l_t \leq 1.1 \,\, m \\
	\end{equation}
         \begin{equation}
         \label{derivativeConst2}
		u_{safe} - u_{t-1} \leq 360 \,\, m^3/s \,\,\,\, if \,\, l_t > 1.1 \,\, m \\
	\end{equation}
	\begin{equation}
			\textbf{s} \in R^H, \textbf{u} \in R^H, \epsilon \in R^H
	\end{equation}
        \begin{equation}
        \label{initiConst}
			u_1 = u_{safe} 
	\end{equation}
 \begin{equation}
        \label{dosConstant}
			u_2 = ... =  u_h = ... = u_{DoS} \,\,\, \forall \,\, 2 \leq h \leq H
	\end{equation}
\end{subequations}
The structure is similar to the \textbf{ResMPC}, but there are substantial differences: 
\begin{itemize}
    \item The optimally computed control action is exclusively the one calculated for the next time instant, as Equation \eqref{initiConst} specifies. The controller is designed assuming that from the following time instant, a DoS attack might occur. If it happens, the controller will lose the link with the corresponding actuator responsible for the execution of the computed control action. The system will go open loop until the DoS attack finishes.  
    \item The controller assumes that the corrupted control action is equal to $u_{DoS}$ and that remains constant for the whole prediction horizon after the current instant. 
\end{itemize}
The other equations follow the same structure of the \textbf{ResMPC}. The optimal control action $u_{safe}$ is computed minimizing the cost obtained after the evolution of the system trajectory as a consequence of the corrupted actions. \textbf{Remark:} we assume reasonably that we can anticipate which is the value of the malicious control action $u_{DoS}$. This assumption derives from the fact that, we assume the attacker will hit with the purpose of causing the highest possible damages, thus applying exactly $u_{DoS}$ whose computation is described in the next section.

Both the ResMPC and the SafeMPC optimization problems are solved using the convex optimization solver CVX. See \cite{cvx}.

\subsection{Optimal Attacker}
\label{OptimalAttacker}
The attacker solves an "adversary" optimization problem at each time instant. Given the same information on the system (states and inputs), it calculates the optimal control action, which, if applied at that instant on the system, maximizes the objective function (which in the MPC optimization problem was instead minimized). We note that the solved problem is no longer convex but rather concave (maximizing a convex cost function is equivalent to minimizing a concave one). This constrains the selection of solvers for the optimization problem. We choose to replace CVX (used in MPC controllers) with Gurobi (\cite{gurobi}), a commercial solver capable of dealing with non-convex functions.

\begin{subequations}
	\label{attackerProblem}
	\begin{equation}
            \label{costAttacker}
		\operatorname{max}_{u_{DoS}}\{c_f||\textbf{s}-s_{max}||_2 + c_d||\textbf{s}-s_{min}||_2+c||\epsilon||_2\}
	\end{equation}    
	\begin{equation}
        \label{massBalance}
		s_{h+1} = s_h +3600\cdot 24(\hat{q_h} - u_h) 
	\end{equation}
	\begin{equation}
        \label{controlConstraintAttacker}
		0 \leq u_{DoS} \leq u_{max}
	\end{equation}
     \begin{equation}
     \label{controlConstraint2}
		u_{DoS} \geq d_h +\epsilon_h \\
     \end{equation}
      \begin{equation}
			\textbf{s} \in R^H, u_{DoS} \in R^H, \epsilon \in R^H
       \end{equation}
\end{subequations}
The Equations \eqref{attackerProblem} describe the constrained concave problem solved at each time step by the attacker. We highlight the maximization problem in Equation \eqref{costAttacker}. We also highlight the difference (with respect to Equations \eqref{controlConstrainta} and \eqref{controlConstraintSafe}) in the lower bound in Equation \eqref{controlConstraintAttacker} on attacker control action which is $0$ $m^3/s$ and not $u_{min}$ (as it is the value given by the required minimum environmental flow (see \cite{cestari2022hourly}), not respected by the attacker.).
We assume the attack duration is fixed and equal to $1$ day.

\begin{figure}[htbp!]
\centering
\includegraphics[width=1.02\columnwidth]{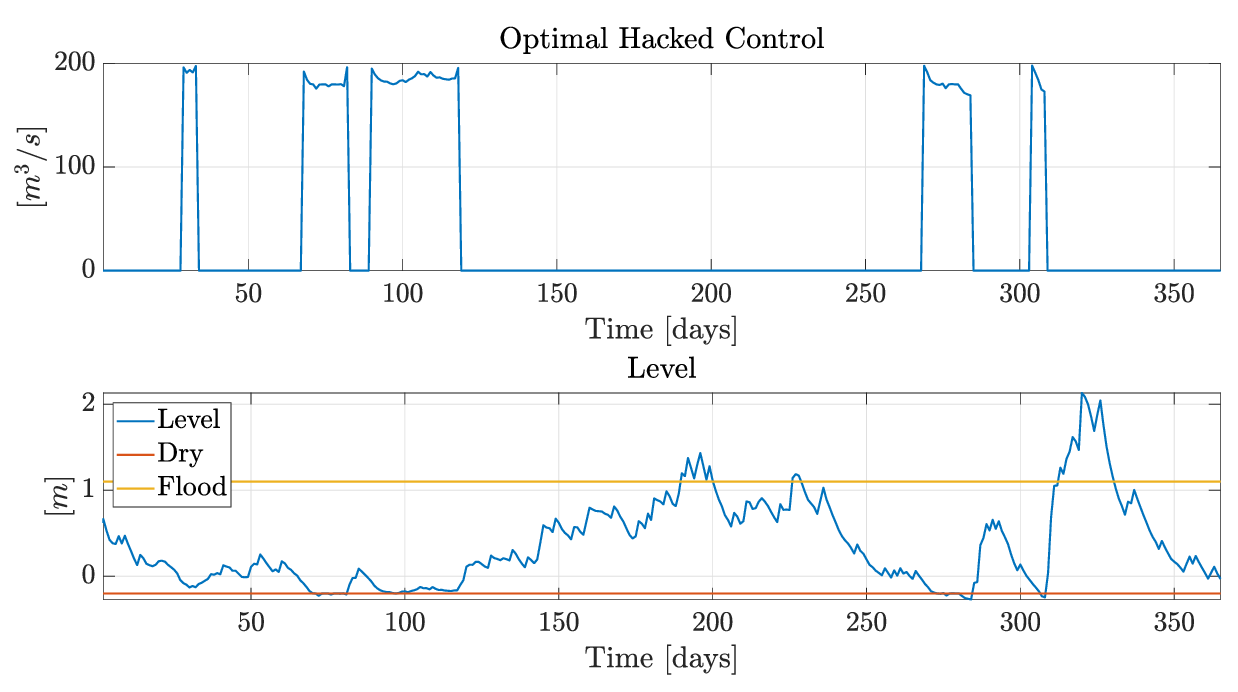}
\caption{\textit{Maximum Damage} control (upper panel), Lake Como level (lower panel)}
\label{fig:hacked}
\end{figure}

Figure \ref{fig:hacked} shows in the lower panel the Lake Como level and in the upper panel the correspondingly computed optimal "maximum damage" control action, obtained as solution of Equations \eqref{attackerProblem}. In the dry periods, the best choice for the attacker is applying the maximum water release (compliant with the upper bound, given by the dam physical structure), wasting the resource. In flood periods, the best choice is the opposite, closing the dam and amplyifing the flood effect. The optimal attack adapts online, monitoring the lake condition at each time step.

\subsection{Adaptive Resilience Motivations}
\label{motivations}
\begin{figure}[h!]
\centering
\includegraphics[width=1.02\columnwidth]{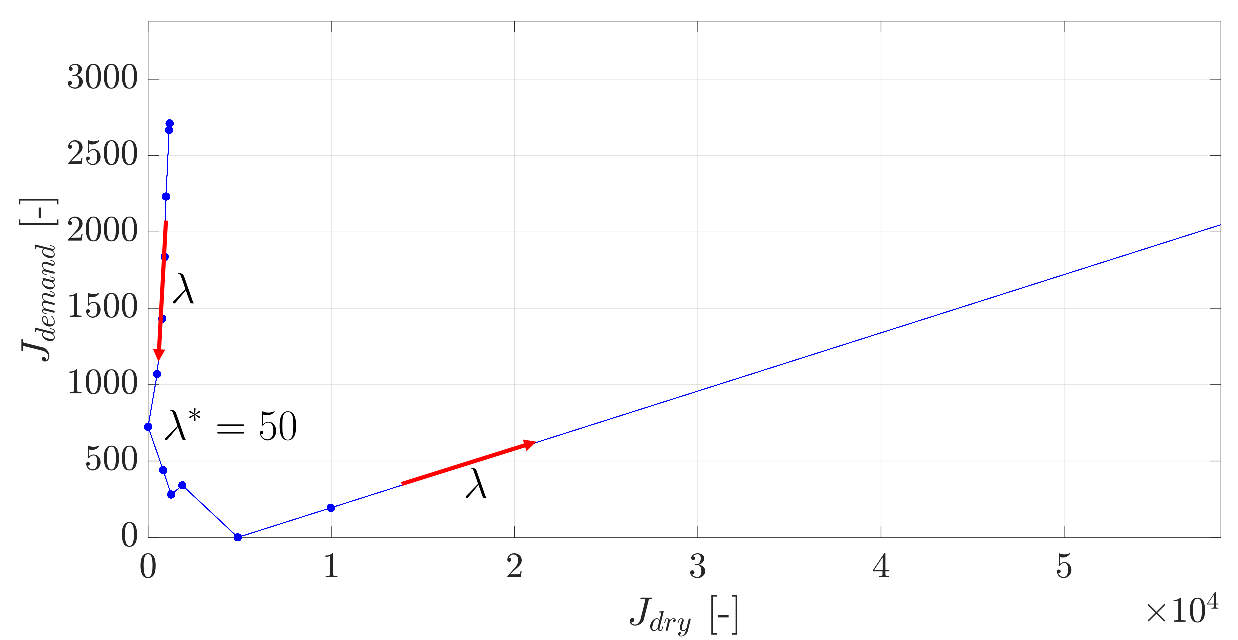}
\caption{$\Bar{\lambda}$ sensitivity analysis for fixed resilience, Resilient MPC.}
\label{fig:sensitivity}
\end{figure}

In this section, we illustrate why adopting a resilience factor $\lambda$ that can adapt online is necessary. Given how much we want to weigh the conservative control action of "safety," the correct level of resilience depends on the attack condition. If the attack frequency is high, a conservative solution is best. However, in sporadic attacks, the performance-oriented solution is convenient.
Table \ref{tab: fixedResilienceTable} shows the control cost variation ($\Delta J$ [-]) with respect to the minimum, between the MPC, Resilient MPC (with fixed resilience factor $\lambda=\Bar{\lambda})$ and Safe MPC control strategies, for different attack frequencies. The results confirm the previous discussion. Resilient MPC outperforms the competitors only in a "in-between scenario" where the attack frequency is of 1 attack every 2 weeks. This motivates the need of resorting to an adaptive resilience factor $\lambda=\lambda(t)$. 
\textbf{Remark:} $\Bar{\lambda}=\lambda^*=50$ is computed after sensitivity analysis, finding the best $\Bar{\lambda}$ minimizing the Resilient MPC control cost, as Figure \ref{fig:sensitivity} shows as best trade off solution for the Pareto front.
\textbf{Remark:} Control cost $J$ is the MPC control cost described in Equation \eqref{cost1}, without the resilience term (e.g., $J = c_f||\textbf{s}-s_{max}||_2 + c_d||\textbf{s}-s_{min}||_2+c||\epsilon||_2$) evaluated a posteriori on system trajectories, for each control strategy.
\textbf{Remark:} $J_{demand} = c||u-d||_2$, $J_{dry} = c_d||\textbf{s}-s_{min}||_2$ are the dominating cost components used in Figure \ref{fig:sensitivity}. Flood cost component is equal for all the tested $\Bar{\lambda}$ hence is not shown.

\begin{table}
\centering
\begin{tabular}{ |c|c|c|c| } 
\hline
$\Delta J$ [-] & \textbf{MPC} & \textbf{Resilient MPC} & \textbf{Safe MPC}\\
 \hline
\textbf{ No attacks }& $0$ & $4943$ & $41845$ \\
 \hline
 \textbf{1 attack per week }& $14645$ & $12516$ & $0$  \\
 \hline
 \textbf{1 attack every 2 weeks} & $1488$ & $0$ & $9853$ \\
 \hline
\end{tabular}
 \caption{ $\Delta$ Control Cost - MPC, Resilient MPC, Safe MPC - No attacks, 1 attack per week, 1 attack every 2 weeks.}
\label{tab: fixedResilienceTable}
\end{table}

\subsection{Hawkes-based Adaptive Resilience}
\label{hawkes}
In this section we show how we derived a time-varying resilience factor adapting to varying attack conditions. At each time step, we train a Hawkes process, we compute the estimated intensity function for different attack return times and at last, we integrate the intensity function to retrieve the resilience factor. The higher the probability of having attacks shortly, the higher will be the resilience factor.

Each data (attack) vector on which the process is trained is a binary vector ($1$ if attack occurs, $0$ otherwise) of length equal to the simulated time steps up to the current time. At each step, the attack vector is updated and stored into database.

\begin{figure}[h!]
\centering
\includegraphics[width=0.9\columnwidth]{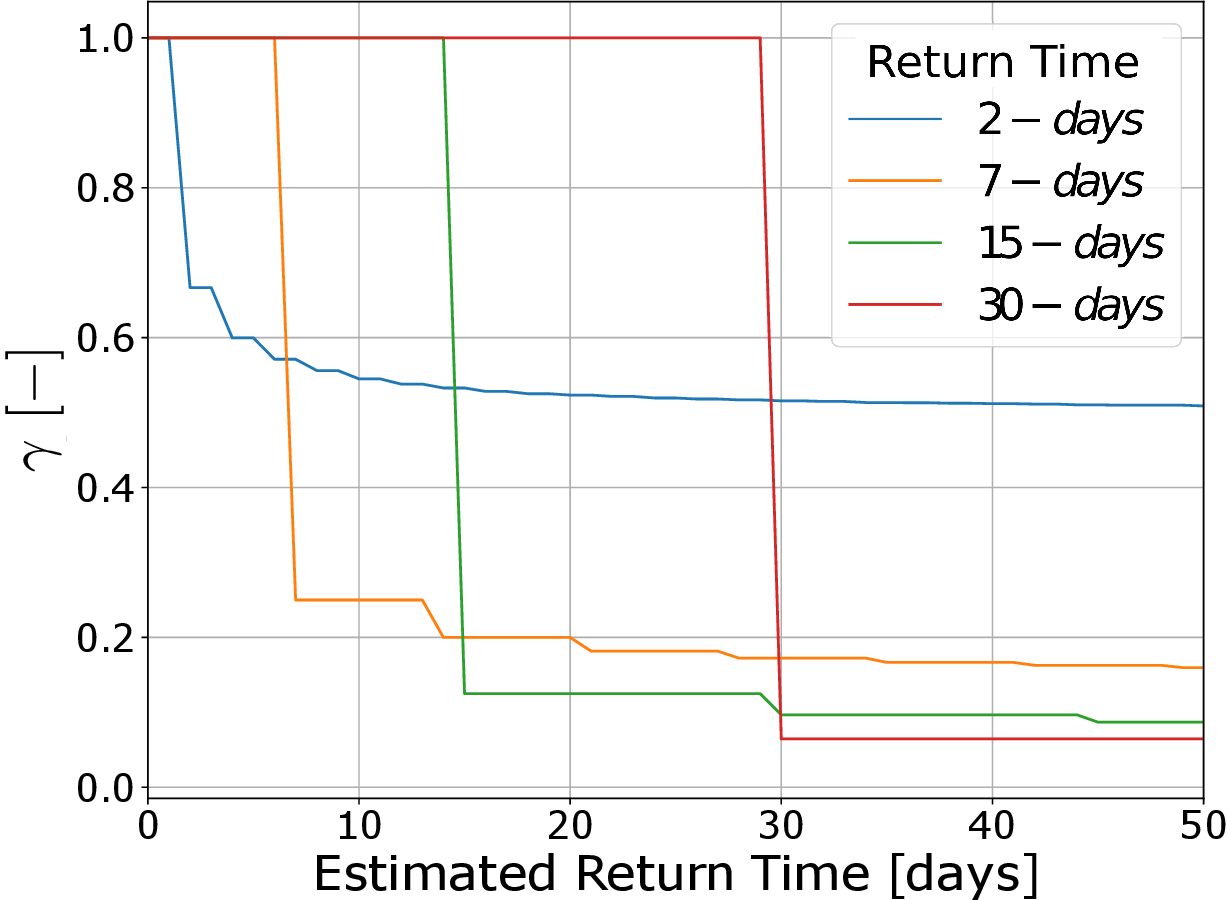}
\caption{Hawkes intensity function as function of return time, for different attack return times.}
\label{fig:intensity}
\end{figure}

Hawkes processes are a class of point processes. The Hawkes choice depends on their wide application in predicting the occurrence of random events in time. In Hawkes, the intensity function is defined (according to \cite{hawkes1971spectra}) as:
\begin{equation}
\label{intensity}
\gamma(t) = \mu + \sum_{k}^{k(t_k<t)}\alpha \cdot e^{-\beta \cdot (t - t_k)},
\end{equation}
where $t$ is current time, $t_k$ is the event time (in our case, attack time), $\mu$ is the baseline, (e.g., the event rate in the absence of past events), $\alpha$ is the weight associated with each event, $\beta$ is the decay rate and controls the exponential decay over time. 
The collection $\theta = [\mu, \alpha, \beta]$ constitutes the parameter vector that must be identified during Hawkes training.
$\theta^*=[\mu^*, \alpha^*, \beta^*]$ is identified through maximum likelihood estimation (MLE) on a moving window $W=W^*$ of historical data.
MLE identifies $\theta^*$ maximizing the joint probability of the observed data, e.g., the likelihood function $\mathcal{L}_n(\theta)$,
\begin{equation}
\label{likelihood}
    \theta^* = \underset{\theta \in \Theta}{\operatorname{arg\;max}}\, \mathcal{L}_n(\theta ; \mathbf{y}),
\end{equation}
where $\textbf{y}$ is the collection of observed data samples and $\Theta$ is the parameter space.
The window size affects the quality of predicted intensity function, however: 
\begin{itemize}
    \item a larger window mandates a higher warm-up time in which the attack detection process is not active.
    \item a larger window slows down the adaptive capacity of the Hawkes process to sudden attack pattern changes in time.
    \item a shorter window leads to less data available for Hawkes process training, increasing the probability of overfitting.
\end{itemize}
For the above reasons, our choice is a window size of $W^*=30$, fulfilling the rule of thumb of a ratio of $10$ between the data numerousness and the number of parameters, guaranteeing overfitting avoidance, and at the same time, minimizing the warm-up time.
After Hawkes process training, we can compute the intensity function for different event (attack) return times. The higher will be the intensity function on the corresponding horizon, the higher the probability of having an attack within the given return time. Figure \ref{fig:intensity} shows the Hawkes intensity function (after training) for different estimated return times, under different attack vectors (with different attack return periods). The intensity function stays at its maximum for the length of the expected return time, then it decreases. 
At each time step, the intensity function is "discretely integrated" to retrieve the resilience factor.
\vspace{-0.2cm}
\begin{equation}
\label{resF}
    \lambda(t) = \sum_l^{L_{MAX}} \gamma_l(t)
\end{equation}
where $l$ is the varying return time and $L_{MAX} = 100$ is the maximum return time considered. The choice of this hyperparameter does not affect Hawkes prediction quality as it represents exclusively the number of lags considered for the intensity function evaluation. The resilience factor (e.g., intensity function integral) is higher as the estimated attack return time is shorter. The Hawkes process fitting procedure is performed in an expanding window fashion on simulation data if the current simulation time is lower than the training window size (hence, the warm-up period is required). When the simulation time is equal to or greater than the training size, only the last $W^*$ elements of the window (including the current one) are read from database and used for training. For perspective at a glance see the flow chart in Figure \ref{fig:scheme} (bottom left for the attack detection components). The algorithm is implemented in Python, thanks to the Machine Learning open-source library Tick, which implements the Hawkes training procedure. See \cite{bacry2018tick}.
\begin{algorithm}[H]
	\caption{Adaptive Resilient MPC Algorithm}
	\label{alg:MYALG}
	\begin{algorithmic}[1]	
		%\Procedure{Roy}{$a,b$}       %\Comment{This is a test}
		\For{t = 1,...,$T_{sim}$}
            \State Solve \textbf{Attacker Problem }with \eqref{attackerProblem}, return $u_{DoS}$.
            \State Train \textbf{Hawkes Process} with \eqref{intensity}, return $\gamma$.
            \State Compute the resilience factor $\lambda$ with \eqref{resF}.
            \State Solve \textbf{SafeMPC} with \eqref{SafeMPC}, return $u_{safe}$.
		\State Solve \textbf{ResMPC} with \eqref{MPC}, return $u_{res}$.
            \State \textbf{If} attack occurs, apply $u_{DoS}$ to the system.
            \State \textbf{Else}, apply $u_{res}$.
            \State Update attack occurrence database.
        \EndFor
        \textbf{End for.}
		%\EndProcedure
	\end{algorithmic}
\end{algorithm}

Algorithm \ref{alg:MYALG} describes the Adaptive Resilient Model Predictive Control procedure. $T_{sim}$ is the simulation length, in our case of $1$ year ($365$ days).

\vspace{-0.32cm}

\section{Numerical Simulations and Results}
\label{results}
In this section we show the numerical simulations and results obtained with the 3 compared control strategies: MPC, Adaptive Resilient MPC, and Safe MPC in 2 different attack scenarios.

In the first case, we consider a time-varying periodic attack return time. In the second we tackle an attack occurrence prescribed by a pseudo random binary signal (PRBS). 

We perform the numerical simulations with real data of year 2005. All computations are carried out on an Intel Core i7-8750H with 6 cores, at 2.20 GHz (maximum single core frequency: 4.10 GHz), with 16 GB RAM and running Matlab R2021b. The control system runs in Matlab-Simulink, the MPC optimization steps are implemented through the CVX solver. The optimal attacker runs in Python 3.7 through the Gurobi solver (Gurobipy extension). Hawkes algorithm runs in Python 3.7 through Tick statistical learning library.

The first attack scenario is a time varying periodic attack defined by the following piece-wise conditions. This scenario is relatively easy to detect but still, is representative for the effectiveness of the introduction of the degree of freedom given by the time-varying resilience factor, and, simultaneously, the capability of Hawkes process to adapt to changing patterns.

\begin{equation}
\label{timeVaryingAttackPatter}
T = \begin{cases}
7 & \text{if } t < 50 \text{ or } t > 250 \\
14 & \text{if } 50 \leq t < 150 \\
21 & \text{if } 150 \leq t \leq 250
\end{cases}
\end{equation}

Equations \eqref{timeVaryingAttackPatter} prescribe a periodic attack pattern with changing return period $T$ of $1$, $2$, and $3$ weeks. 
\textbf{Remark:} simulation horizon is of $1$ year ($365$ days), daily resolution.

The second attack scenario is defined by a PRBS (pseudo random binary signal) with clock time $\tau=10$ (e.g., the number of instants for which the signal must stay constant) with the assumption of not having attacks in consecutive days. This scenario is closer to reality as the attacker would like to be as unpredictable as possible in the definition of the attack times. The choice of a PRBS as triggering signal allows us to define a random signal with still a pattern that could be tracked by the Hawkes process, given by the choice of the clock period.

\begin{figure}[h!]
\centering
\includegraphics[width=1.02\columnwidth]{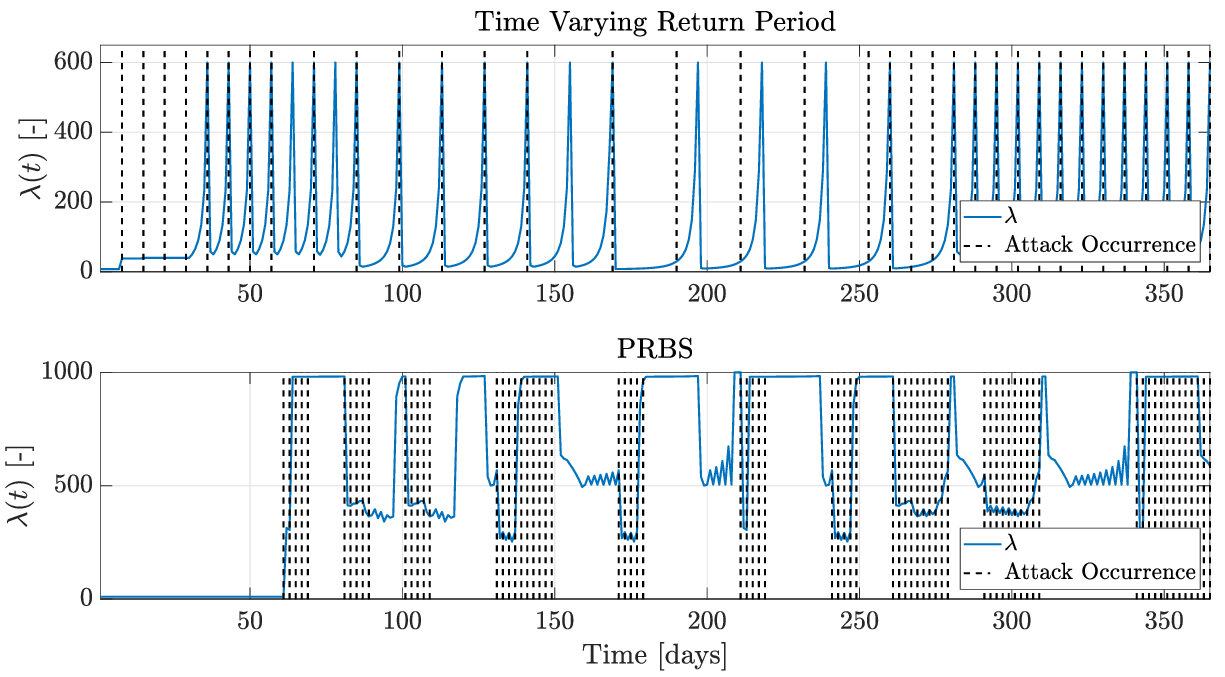}
\caption{Resilience Factor $\lambda$ and attack occurrence.}
\label{fig:factor-varying}
\end{figure}

Figure \ref{fig:factor-varying} shows the time-varying resilience factor obtained through Equations \eqref{intensity} and \eqref{resF} in time-varying return period scenario (upper panel) and in PRBS scenario (lower panel). In the upper panel, we highlight the required warm-up window $W$ in which even when the first attack occurs the resilience factor increases but takes a whole window to detect the first pattern ($7$ days). When $T$ return period changes the process tracks the changing pattern with a delay. This is the proof that, the Hawkes process can detect time-varying periodic patterns. 

Figure \ref{fig:factor-varying}, lower panel, highlights how the resilience factor adapts also in case of pseudo-random signal occurrence. Hawkes as soon as detects the occurrence of the attacks increases the resilience level leading to a more conservative control action. However, it modulates the resilience on the estimated attack return period "charging and discarging the integral" given by Equation \eqref{resF}. 

As Table \ref{tab: adaptiveResilienceTable} shows, in both the scenarios the proposed ARMPC architecture outperforms the competitors "performance-oriented" and "safety-oriented" strategies. Cost margin is higher in the time-varying return period scenario as Hawkes detects with accuracy the changing return time of the attack. In the pseudo-random case the advantage is still evident even if with lower margin.

\begin{table}
\centering
\begin{tabular}{ |c|c|c|c| } 
\hline
$\Delta J$ Adaptive Resilient MPC [-] & \textbf{MPC} & \textbf{Safe MPC}\\
 \hline
\textbf{Time-varying Return Period}& $6720$ & $1349$ \\
 \hline
 \textbf{PRBS}& $7532$ & $119$  \\
 \hline
\end{tabular}
 \caption{ $\Delta$ Control Cost with respect to Adaptive Resilient MPC - MPC, Safe MPC - Time-varying Return Period, PRBS.}
\label{tab: adaptiveResilienceTable}
\end{table}

\begin{figure}[h!]
\centering
\includegraphics[width=\columnwidth]{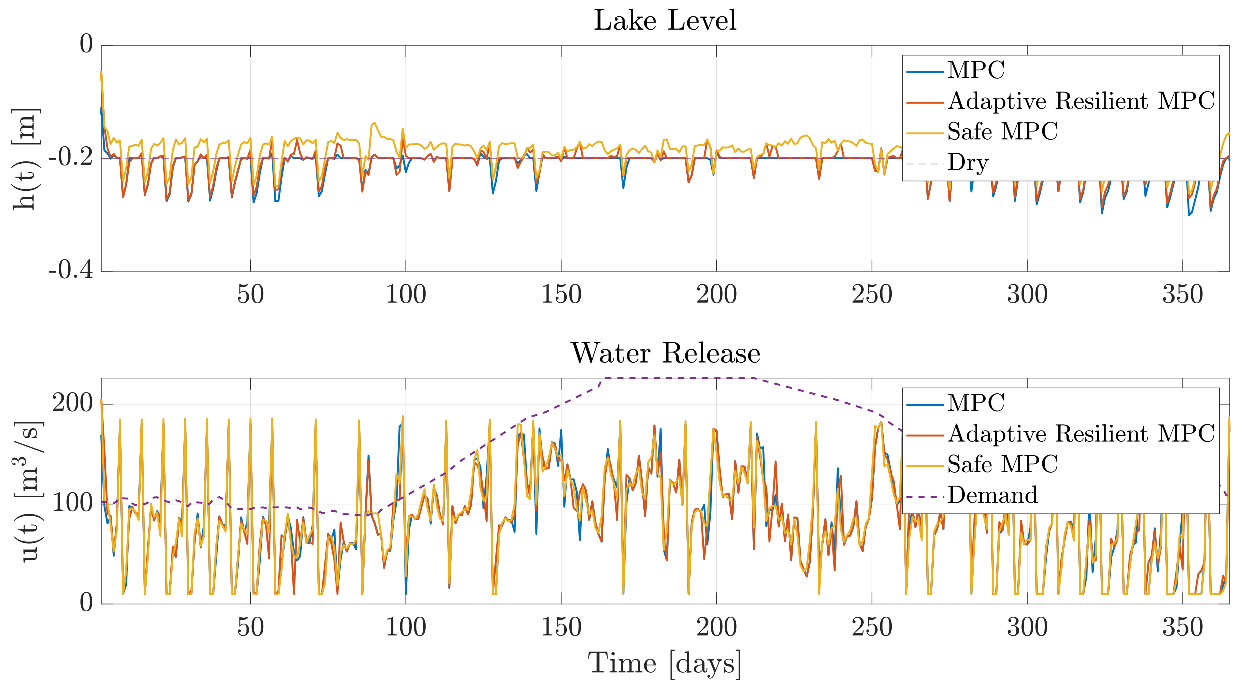} % level-adaptive-integral.eps
\caption{Upper panel: Lake Como level, lower panel: water release - 2005 - Time-varying Attack Return Period.}
\label{fig:level}
\end{figure}

Figure \ref{fig:level} shows the Lake Como level in 2005 (upper panel) and water release (lower panel) with the 3 control strategies under optimal attack with time-varying return period attack pattern. SafeMPC is conservative. It always increases lake level (reducing water release). Standard "performance-oriented" MPC works as if attacks do not happen. Adaptive Resilient MPC instead, estimates the attack occurrence time and increases resilience level as the return time gets closer. This is evident in correspondence of correctly estimated return times where the lake level at first sticks to the lower bound, then increases, for the resilience "activation" and then decreases, given the occurrence of the attack. In this way, it avoids lake level negative peaks and, at the same time, tracks at best the agricultural demand. Under attack conditions, all the strategies see $u_{DoS}$ control action (as the attack "spikes" underline in Figure \ref{fig:level}, lower panel), given the current system condition.

\section{CONCLUSIONS}
\label{conclusions}
The proposed adaptive resilience architecture (ARMPC) allows us to maximize the performance of an MPC control scheme even in the presence of DoS attacks, which inhibit the policy the controller decides and replace it with an optimal attack, maximizing system damages. The use of Hawkes processes to estimate the return time of the attack and the linking of its intensity function (estimation of the probability of event occurrence) with the resilience factor introduces a degree of freedom to the architecture that allows us to achieve the same level of performance in normal conditions and to robustify the control policy in attack conditions. We demonstrate the approach's validity with the attack's occurrence given by a periodic return time with a variant period and given by a pseudo-random signal. In future work, we will use the same architecture in different applications, even on smaller-scale hardware systems. We will also extensively validate the strategy over multiple years of real system data.

\addtolength{\textheight}{-12cm}   % This command serves to balance the column lengths

\bibliographystyle{IEEEtran}
\bibliography{bibliography}

\end{document}